\documentstyle[aps,pre,multicol,epsfig]{revtex}

\newcommand{\rhog}{\mbox{$\phi$}}
\newcommand{\rhoe}{\mbox{$\rho$}}
\newcommand{\uf}{\mbox{$u$}}

\newcommand{\as}{\mbox{$\epsilon$}}

\begin{document}

\title{Formation of Polymer Brushes}
\author{J. Wittmer$^1$ and A. Johner$^2$}
\address{
$^1$ Dept. of Physics and Astronomy, University of Edinburgh\\
JCMB King's Buildings, Mayfield Road, Edinburgh EH9 3JZ, UK.}
\address{$^2$ Institut Charles Sadron (UPR CNRS 022)\\
6 rue Boussingault, 67083 Strasbourg Cedex, France}

\date{July 1996}

\maketitle

\begin{abstract}
In systems such as block copolymer mesophases or physical gels formed by 
associating  copolymers, the dynamical properties are often controlled by 
the extraction/association of a sticking group. We propose a description 
of the extraction/association process of a single sticker. The statistical
physics of these associated systems is usually dominated by stretched 
brush-like regions. A sticker has to overcome a potential barrier both 
to penetrate the stretched structure or to escape a favorable region built 
by associated stickers. Our main result is that these barriers are crossed by 
tension fluctuations and that the corresponding processes are thus local with 
a friction independent of molecular weight. When the potential barriers are 
high, the (very stretched) equilibrium structures are not likely to develop
on reasonable time scales. Stretched model systems may also be  grown {\it in 
situ} from nuclei bearing initiating groups. These, irreversibly bound 
structures, are also briefly discussed.  
\end{abstract}

\begin{multicols}{2}

Molten block copolymers self-assemble forming various structures mainly 
depending on their asymmetry \cite{ludwik80,sashablocks}.
In a selective solvent, say good for $A$ bad for $B$, the $A$-blocks are
swollen by the solvent whereas the $B$-blocks assemble in almost solvent 
free domains \cite{carlos}. Soluble polymers decorated with insoluble
 stickers form a physical gel where the temporary crosslinks are build by 
aggregated stickers \cite{sashatriblocks}. For such a material to flow, 
stickers have to be extracted from aggregates. There is usually a high energy 
barrier $E$ to overcome during the extraction process. In the case where 
the stickers are small insoluble $B$-blocks $E\sim \gamma N_B^{2/3}$ and 
a high tension $\tau$ of order $a\gamma$ with $a$ the monomer size and $\gamma$
the $B/solvent$ surface tension is needed for non activated chain extraction.
Thermally activated chain extraction (or sticker desorption) is thus an 
important issue. Long soluble blocks strongly interact in the vicinity 
of an insoluble domain and stretch to avoid each other\cite{ADG}. Even in the 
case of associating polymers forming a physical gel, it is believed that the 
star like regions  around  small insoluble domains dominate the statistical 
physics of the network\cite{duplantier1}. 
The extraction of a sticker belonging to a locally stretched chain is thus 
of rather general relevance.

In the early work, over the past decade, the random motion of the sticking 
group has been described as that of a point-like particle
with the friction $N\zeta$ relevant for the overall motion of the chain. 
Here we argue that the internal modes of the chain are important an that 
the relevant friction is much lower, it rather corresponds 
to the first correlation length\cite{PGGb} $\xi$, linked to the aggregate 
parameters through $p\xi^{d-1}\sim S$ where $p$ is the functionality of the 
aggregate and $S$ its area\cite{daoudcotton} (for insoluble blocks 
$S\sim(pN_B)^{2/3}$). The physical argument is as follows:
once the had group is off the aggregate it almost freely diffuses over the 
first correlation length $\xi$ and bounces back on the aggregate many times, 
 at distances larger than $\xi$, the equilibrium tension 
$\tau_e\sim k_bT/\xi$ drives the motion of the sticker and extraction is 
achieved. This is to say that extraction is a local process independent of 
the overall chain length $N$.

\begin{figure}[tbh]
\centerline{\epsfysize=6cm
\epsfig{file=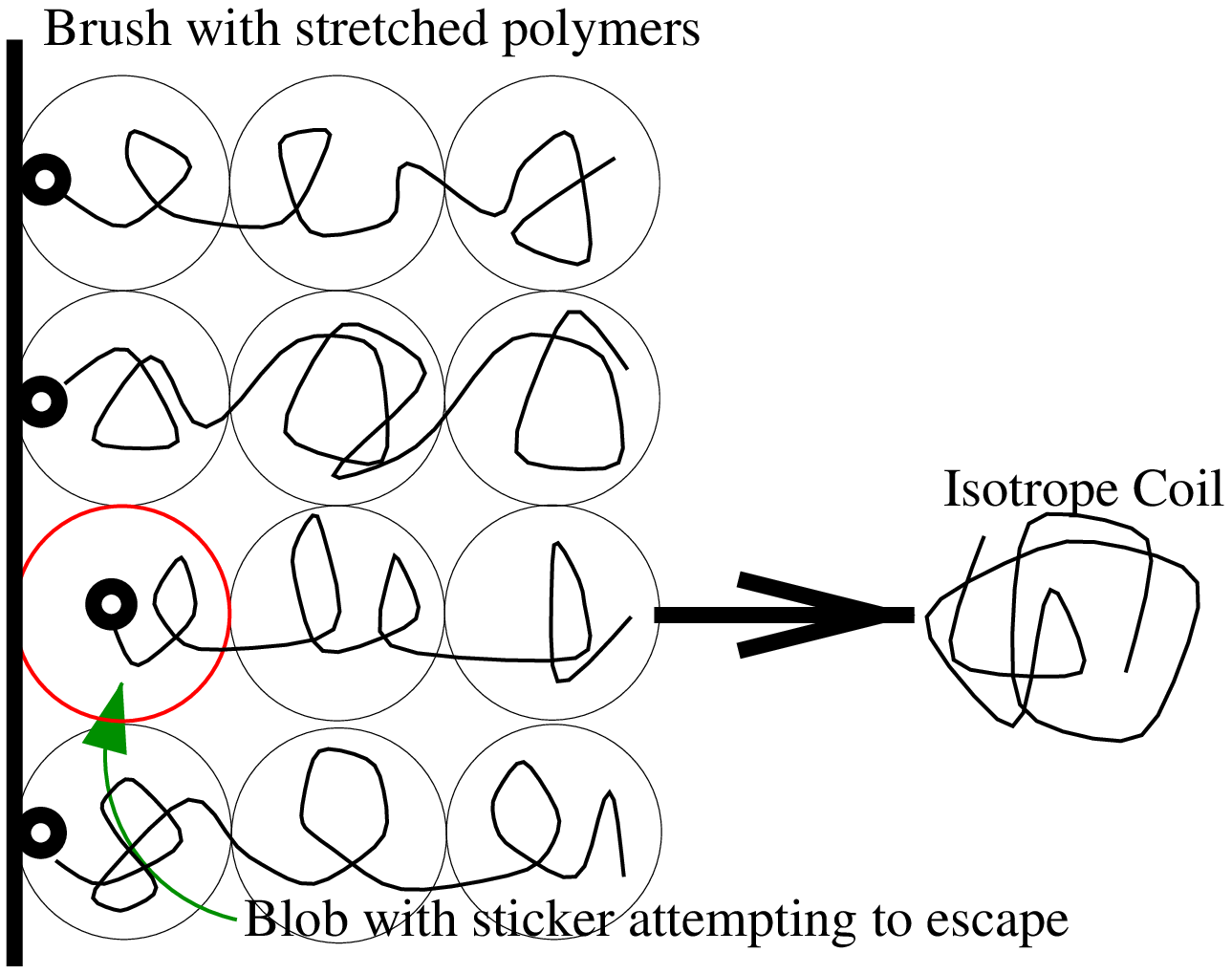,width=60mm,height=50mm,angle=0}}
{\small FIG.~1. Sketch of successful desorption attempt of a 
polymer chain out of a semi-dilute polymer brush.
\label{fig:desorb}}
\end{figure}

The idea that the relevant friction should be local can be tested on the 
simpler case where a chain is cut off the grafting surface in a polymer brush 
(with no sticker left).
Due to the retraction of the chain there is a drift of the chain end.
The relevant scales are the height $h$ of the brush and the relaxation time of 
the typical stretched configuration $\tau_r$. 
The following scaling form must thus hold for the motion of the cut end:
\begin{equation}
\left<z\right> = h{\rm f}(t/\tau_r) \quad {\rm whith:}\,\tau_r\sim 
\left({h\over\xi}\right)^2\xi^d
\label{scalinggen}
\end{equation}
d is the dimension of space, $d=3$ corresponds to excluded volume statistics and
Rouse-Zimm dynamics 
whereas $d=4$ corresponds to mean-field statistics and Rouse dynamics. Chain 
retraction being a local process
 ${\rm f}(x) \sim x^{1/2}\quad (x\ll 1)$ and for $t\ll\tau_r$ :
\begin{equation}
<z>\sim\xi\left({t\over\xi^d}\right)^{1/2}
\label{average}
\end{equation}
The early motion of the chain is nonetheless dominated by anomalous diffusion: 
the chain end excites longer and longer modes whilst moving 
and the friction increases. This leads to the classical dispersion 
law\cite{doib}:
\begin{equation}
<(z-<z>)^2>\sim t^{2/d}
\label{dispersion}
\end{equation}
The motion is thus driven by the tension for $<z>>\xi$ (or $t>\xi^d$) when the 
fluctuation around the average 
position $<z>$ is negligible. It is then unlikely that the chain end hits the 
grafting surface again.
These ideas are supported by both  a Monte Carlo simulation and a Rouse 
analysis\cite{jojo1}.

Similar ideas hold for thermally  activated had group desorption from a flat 
surface. Once the head group is off the grafting 
surface, desorption is promoted by chain retraction and becomes irreversible for
$z>\xi$. The process
being again local the relevant time scale depends on $\xi$. Assuming that at the
scale of the 
correlation length the only time scale
 is the cooperative  relaxation time $\xi^d$ the outwards flux  obeys:
\begin{equation}
J_{out} \sim \xi^{-d} \exp(-E) \sigma \sim \xi^{1-2d}\exp{-E}
\label{jout}
\end{equation}
and is independent of chain length, the grafting density $\sigma$ is linked to 
$\xi$ via 
$\sigma\sim\xi^{1-d}$. The characteristic lifetime of a bound sticker $T_-$ is 
deduced from rate 
equation (\ref{jout}) as:
\begin{equation}
T_- \sim \xi^d\exp E
\end{equation}
This also applies to curved surfaces.
This result is supported by Monte Carlo simulations\cite{jojo1}.

The inward flux $J_{in}$ is limited by the barrier of height $\mu$ opposed to 
sticker penetration by 
already grafted chains. Assuming that the incoming sticker also crosses the 
barrier by a local 
tension fluctuation, $\xi$ is the only relevant scale. The inwards flux thus 
obeys:
\begin{equation}
J_{in} \sim \xi^{1-d} \exp(-\mu) c_b
\label{jin}
\end{equation} 
with $c_b$ the bulk chain concentration. 
Note that the kinetic equations (\ref{jout},\ref{jin}) correspond to the 
isotherm:
\begin{equation}
\mu_{eq} - E \sim \log(c_b/c^\star) -{d\over d-1}\log(\sigma/\sigma^\star)
\label{isotherm}
\end{equation}
where $\sigma^\star\sim R^{1-d}$ and $c^\star\sim R^{-d}$ are the  overlap 
concentrations for 
grafted chains and free chains respectively. The chemical potential increment 
$\mu$ for a grafted chain 
with respect to a free chain is $\mu\sim h/\xi$ to leading order, it also 
contains logarithmic corrections 
involving enhancement exponents.

Due to the activation barrier, the extraction/aggregation process is slow.
 The relaxation time of small 
fluctuations in the aggregation number (inside the peak of the size 
distribution) usually lies in the 
minute range for diblock copolymers\cite{tuzar,johner1}. Large fluctuations not 
conserving the number of aggregates are in contrast 
found to relax extremely slowly\cite{johner1}. Conversely the existence of large
equilibrium aggregates is questionable.
One way to overcome this difficulty is to proceed with concentrated solutions 
where the excluded volume 
is screened and to swell the system in the solvent afterwards\cite{auroy}. It is
however unclear how the system relaxes 
topological constraints upon swelling. For some purposes as  colloid coating 
(coated colloids are 
used in 
filled rubbers) chains can be irreversibly grafted. An effective way to achieve 
fairly high grafting 
densities is to grow the layer {\em in situ} monomer by monomer from a 
functionalized nucleus carrying 
polymerisation initiators. We consider a flat surface densely covered with 
initiators. 
There is some similarity between  needle growth\cite{mikeneedles} governed by 
classical D.L.A
\cite{DLA} without branching and
 polymer growth,
nonetheless  polymer  
chains can relax their configurations, this sets an additional dynamical time 
scale.
We consider low enough  growth rates such that the chain configurations are 
completely relaxed and 
the grown structure is  in 
internal thermodynamic equilibrium at any time, this provides an additional 
relation between the bound monomer concentration 
and the density of end points. 

\begin{figure}[tbh]
\centerline{\epsfysize=6cm
\epsfig{file=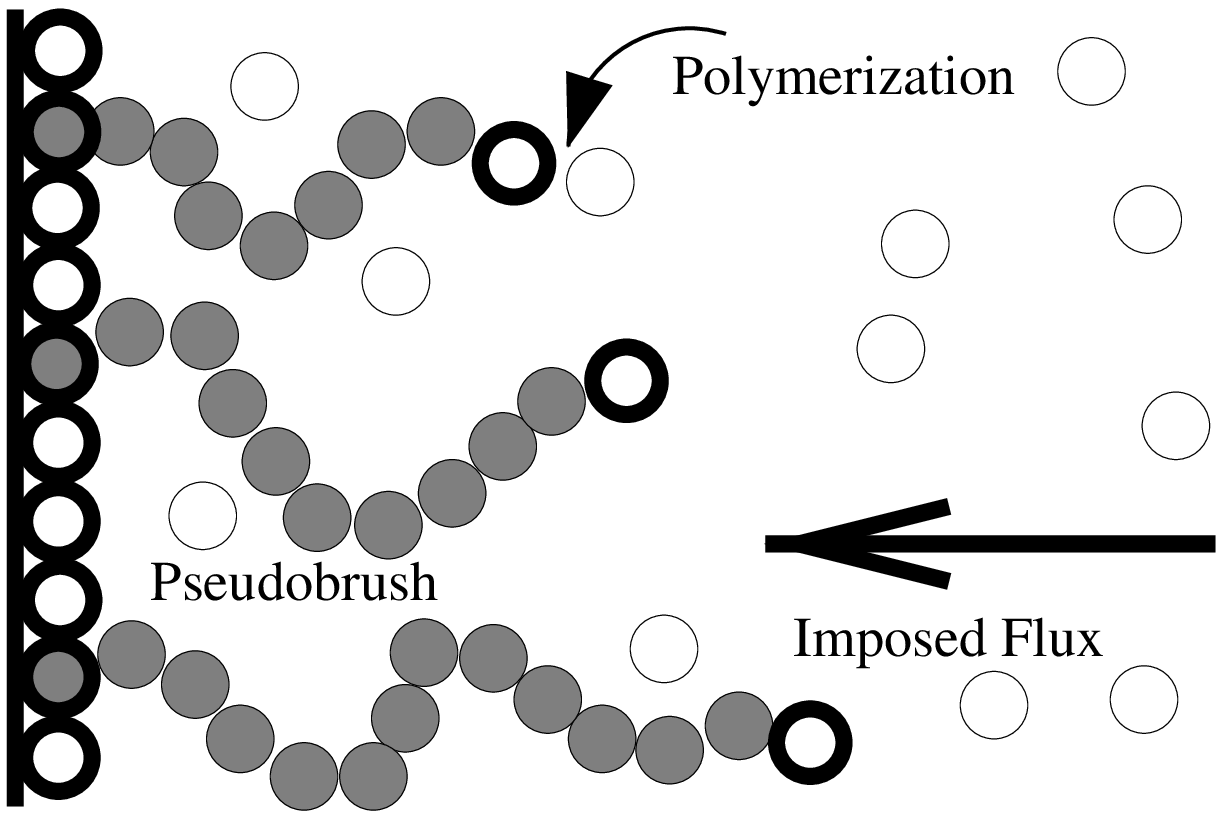,width=60mm,height=40mm,angle=0}}
{\small FIG.~2. Sketch of diffusive growth of a polymer layer by
{\em in situ} polymerization
\label{fig:insitu}}
\end{figure}

In a mean field approach, marginally valid in 3d space\cite{krug}, 
the free monomer density $\uf$, bound monomer density $\rhog$ and 
end point density $\rhoe$ 
are determined by the following set of equations:
\begin{eqnarray}
\rhoe &=& - l \, \partial_{z} \rhog^{1/\as}\nonumber\\
\partial_{t} \rhog &=& k \, \uf \, \rhoe\nonumber\\
\partial_{t} \rhog &=& - \partial_{t} \uf + D \Delta \uf = D \Delta \uf 
\label{dlaeqns}
\end{eqnarray} 
with $l$ a microscopic length of order the monomer size, $\epsilon = 
{d-1/\nu\over d-1}$ an exponent 
linked to equilibrium properties\cite{MWC,ADG} ($\nu$ is the Flory exponent, 
$\epsilon\approx 2/3$), $k$ a kinetic constant and $D$ 
the free monomer diffusion constant. The first equation expresses internal 
equilibrium of the layer 
at any time, the second describes the kinetics of the  polymerisation  reaction 
between chain ends and free 
monomers the third is the monomer conservation law and embodies the {\em 
adiabatic} approximation 
(for a discussion of this point see ref.\cite{jojo2}). This equations are 
supplemented by following boundary conditions:
\begin{equation}
\uf(0,t) = 0\qquad \lim_{z\rightarrow\infty}\,D\partial_z\uf = a^d j_\infty
\label{boundary}
\end{equation}
The density of reacting chain ends is found to (formally) diverge at the wall 
and the flux of free 
monomers can be neglected there $\partial_z\uf (0,t) = 0$. We set $1/k=l=1$ 
thereby choosing the 
time and length units, we further set $D=1$ by an appropriate rescaling of the 
fields $\uf,\,\rhoe,\,
\rhog$.The height of the structure $H(t)$ is defined by the first moment $H(t) =
\int_0^\infty\,z\rhog (z) {\rm d}z/
\int_0^\infty\,\rhog (z) {\rm d}z$. We now seek for a solution of the scaled 
form:
\begin{equation}
\rhog = z^{-\alpha}{\rm f}(x)\quad \rhoe = z^{-\beta}{\rm h}(x)\quad \uf = 
z^{\gamma}{\rm g}(x)
\end{equation}
in the variable $x=z/H(t)$.
The aggregation process eqs.(\ref{dlaeqns},\ref{boundary}) imposes $\beta=2$, 
$\gamma = 1$ and $\alpha = \epsilon$
where the scaling function ${\rm f}(x)$ is assumed to be finite at $0$ and to 
vanish at infinity on 
physical grounds. The height of the brush $H(t)$ increases as a power 
law with time: $H(t)\sim t^{1\over 1-\epsilon}$.
The scaling functions are then determined by solving eqs.(\ref{dlaeqns}) 
numerically. 
In fact the constitutive equation linking the monomer density $\rhog$ and the 
end density $\rhoe$ 
breaks down in the outermost correlation length, our description being coarse 
grained on the scale of 
the local correlation length we have to allow for the function ${\rm f}(x)$ to 
jump to $0$ at the 
brush edge. The  results of this coarse grained analytical mean-field theory are
nicely confirmed by 
Monte Carlo simulations\cite{jojo2}. The grown structure is  densely grafted, 
rather polydisperse
 and
 highly stretched. It should be a good candidate for stabilisation purposes. 
There are early 
 {\em grafting from} experiments\cite{vidal} and very recent ones with a more 
detailed analysis  of the 
obtained structure (mostly unpublished)\cite{ruhe}. The latter use thermally 
controlled radical 
precursors and the initiator formation is rate limiting, the chains are mainly 
grown one by one, 
the  layers seem less 
densely grafted with very long, well stretched, rather monodisperse chains. 
There is hope that {\em in 
situ} growth allows for well stretched layers with various grafting densities, 
mean chain lengths, 
in plane structures 
and polydispersities controlled by the nature (anionic polymerisation has been 
very recently reported)
and the density of the initiators (eventually by temperature or irradiation) and
by the bulk monomer concentration.

\end{multicols}{2}
\end{document}